\pdfoutput=1

\documentclass[runningheads,a4paper]{llncs}

\usepackage[skip=0pt]{caption}
\usepackage{paralist}
\usepackage{enumitem}
\setlist{nosep}

\makeatletter

\renewcommand\section{\vspace*{0.025em}\@startsection{section}{1}{\z@}{-10\p@ \@plus -2\p@ \@minus -4\p@}{2\p@ \@plus 4\p@ \@minus 4\p@}{\normalfont\large\bfseries\boldmath\rightskip=\z@ \@plus 8em\pretolerance=10000 }}

\renewcommand\subsection{\@startsection{subsection}{2}{\z@}{-2\p@ \@plus -4\p@ \@minus -4\p@}{-0.5em \@plus -0.22em \@minus -0.1em}{\normalfont\normalsize\bfseries}}

\makeatother

\usepackage{url}
\usepackage{amsmath}
\usepackage{amssymb}
\usepackage{booktabs}
\usepackage{color}
\usepackage{courier}
\usepackage{enumitem}
\usepackage{graphicx}
\usepackage{float}
\usepackage{footnote}
\usepackage{grffile}
\usepackage{hyperref}
\usepackage[frozencache]{minted}
\usepackage{multirow}
\usepackage[all]{nowidow}
\usepackage[np, autolanguage]{numprint}
\usepackage{setspace}
\usepackage{subfig}
\usepackage{tabularx}
\usepackage{times}
\usepackage{todonotes}
\usepackage[square,comma,numbers,sort&compress,sectionbib]{natbib}
\usepackage{paralist}
\usepackage[labelfont=bf,textfont=bf,skip=0pt]{caption}
\usepackage[ normalem]{ulem}

\usepackage{latexsym}

\usepackage{url}
\urldef{\mails}\path|{cvangysel, e.kanoulas, derijke}@uva.nl|

\renewcommand{\citeauthor}{\citet}

\usepackage{etoolbox}
\AtBeginEnvironment{minted}{\singlespacing%
    \fontsize{8}{8}\selectfont}

\hypersetup{pdfinfo={
Title={Pyndri: a Python Interface to the Indri Search Engine},
Author={Christophe Van Gysel, Evangelos Kanoulas, Maarten de Rijke}
}}

\newminted{python}{mathescape, numbersep=5pt, autogobble, framesep=2mm, fontsize=\footnotesize}

\newfloat{algorithm}{t}{lop}
\floatname{algorithm}{Code snippet}

\makeatletter
\newcolumntype{"}{@{\hskip\tabcolsep\vrule width 1pt\hskip\tabcolsep}}
\makeatother

\renewcommand{\UrlFont}{\ttfamily\scriptsize}

\allowdisplaybreaks

\begin{document}

\mainmatter

\title{Pyndri\thanks{\url{https://github.com/cvangysel/pyndri}}: a Python Interface to the Indri Search Engine}
\author{Christophe Van Gysel \and Evangelos Kanoulas \and Maarten de Rijke}
\institute{University of Amsterdam\\Amsterdam, The Netherlands\\\mails}

\maketitle

\begin{abstract}
We introduce \textbf{pyndri}, a Python interface to the Indri search engine. Pyndri allows to access Indri indexes from Python at two levels:
\begin{inparaenum}[(1)]
	\item dictionary and tokenized document collection,
	\item evaluating queries on the index.
\end{inparaenum}
We hope that with the release of pyndri, we will stimulate reproducible, open and fast-paced IR research.
\end{abstract}

\section{Introduction}

Research in Artificial Intelligence progresses at a rate proportional to the time it takes to implement an idea. Therefore, it is natural for researchers to prefer scripting languages (e.g., Python) over conventional programming languages (e.g., C++) as programs implemented using the latter are often up to three factors longer (in lines of code) and require twice as much time to implement \citep{Prechelt2000}. Python, an interactive scripting language that emphasizes readability, has risen in popularity due to its wide range of scientific libraries (e.g., NumPy), built-in data structures and holistic language design \citep{Koepke2010}.

There is still, however, a lack of an integrated Python library dedicated to Information Retrieval (IR) research. Researchers often implement their own procedures to parse common file formats, perform tokenization, token normalization that encompass the overall task of corpus indexing. \citet{Uysal2014} show that text classification algorithms can perform significantly differently, depending on the level of preprocessing performed. Existing frameworks, such as NLTK \citep{Loper2002NLTK}, are primarily targeted at processing natural language as opposed to retrieving information and do not scale well. At the algorithm level, small implementation differences can have significant differences in retrieval performance due to floating point errors \citep{Goldberg1991fp}. While this is unavoidable due to the fast-paced nature of research, at least for seminal algorithms and models, standardized implementations are needed.
\section{Introducing Pyndri}

Fortunately, the IR community has developed a series of indexing frameworks (e.g., Galago, Lucene, Terrier) that correctly implement a wide range of retrieval models. The Indri search engine \citep{Strohman2005indri} supports complex queries involving \emph{evidence combination} and the ability to specify a wide variety of constraints involving \emph{proximity}, \emph{syntax}, \emph{extracted entities} and \emph{document structure}. Furthermore, the framework has been efficiently implemented using C++ and was designed from the ground up to support \emph{very large databases}, \emph{optimized query execution} and \emph{fast and concurrent indexing}. A large subset of the retrieval models \citep{Zhai2001smoothing,Balog2006experts,Bendersky2010sdm,Guan2013qcm,VanGysel2016sessions} introduced over the course of history can be succinctly formulated as an Indri query. However, to do so in an automated manner, up until now researchers were required to resort to C++, Java or shell scripting. C++ and Java, while excellent for production-style systems, are slow and inflexible for the fast prototyping paradigm used in research. Shell scripting fits better in the research paradigm, but offers poor string processing functionality and can be error-prone. Besides, shell scripting is unsuited if one wants to evaluate a large number of complex queries or wishes to extract documents from the repository as this incurs overhead, causing avoidable slow execution. Existing Python libraries for indexing and searching, such as PyLucene, Whoosh or ElasticSearch, do not support the rich Indri language and functionality required for rapid prototyping.

\begin{algorithm}
\begin{minted}[frame=lines]{python}
index = pyndri.Index('/opt/local/clueweb09')

for int_doc_id in range(index.document_base(),
                        index.maximum_document()):
    ext_doc_id, doc_tokens = index.document(int_doc_id)
\end{minted}
\caption{Tokenized documents in the index can be iterated over. The \texttt{ext\_doc\_id} variable in the inner loop will equal the document identifier (e.g.,  \texttt{clueweb09-en0039-05-00000}), while the \texttt{doc\_tokens} points to a tuple of integers that correspond to the document term identifiers.\label{example:iteration}}
\end{algorithm}

We fill this gap by introducing pyndri, a lightweight interface to the Indri search engine. Pyndri offers read-only access at two levels in a given Indri index.

\subsection{Low-level access to document repository}
First of all, pyndri allows the retrieval of tokenized documents stored in the index repository. This allows researchers to avoid implementing their own format parsing as Indri supports all major formats used in IR, such as the trectext, trecweb, XML documents and Web ARChive (WARC) formats. Furthermore, standardized tokenization and normalization of texts is performed by Indri and is no longer a burden to the researcher. Code snippet~\ref{example:iteration} shows how a researcher can easily access documents in the index. Lookup of internal document identifiers given their external name is provided by the \texttt{Index.document\_ids} function.

\begin{algorithm}
\begin{minted}[frame=lines]{python}
index = pyndri.Index('/opt/local/clueweb09')
dictionary = pyndri.extract_dictionary(index)

_, int_doc_id = index.document_ids(
    ['clueweb09-en0039-05-00000'])
print([dictionary[token_id]
       for token_id in index.document(int_doc_id)[1]])
\end{minted}
\caption{A specific document is retrieved by its external document identifier. The index dictionary can be queried as well. In the above example, a list of token strings corresponding to the document's contents will be printed to \texttt{stdout}.\label{example:retrieval}}
\end{algorithm}

The \texttt{dictionary} of the index (Code snippet~\ref{example:retrieval}) can be accessed from Python as well. Beyond bi-directional token-to-identifier translation, the dictionary contains corpus statistics such as term and document frequencies as well. The combination of index iteration and dictionary interfacing integrates conveniently with the Gensim\footnote{\url{https://radimrehurek.com/gensim}} package, a collection of topic and latent semantic models such as LSI \citep{Deerwester1990lsi} and word2vec \citep{Mikolov2013word2vec}. In particular for word2vec, this allows for the training of word embeddings on a corpus while avoiding the tokenization mismatch between the index and word2vec. In addition to tokenized documents, pyndri also supports retrieving various corpus statistics such as document length and corpus term frequency.

\begin{algorithm}
\begin{minted}[frame=lines]{python}
index = pyndri.Index('/opt/local/clueweb09')

for int_doc, score in index.query('obama family tree'):
    # Do stuff with the document.
\end{minted}
\caption{Simple queries can be fired using a simple interface. Here we query the index for topic \texttt{wt09-1} from the TREC 2009 Web Track using the Indri defaults (Query Language Model (QLM) with Dirichlet smoothing, $\mu = 2500$).\label{example:simple_query}}
\end{algorithm}

\subsection{Querying Indri from Python}
Secondly, pyndri allows the execution of Indri queries using the index. Code snippet~\ref{example:simple_query} shows how one would query an index using a topic from the TREC 2009 Web Track using the Indri default retrieval model.
\begin{algorithm}
\begin{minted}[frame=lines]{python}
index = pyndri.Index('/opt/local/clueweb09')
query_env = pyndri.QueryEnvironment(
    index, rules=('method:dirichlet,mu:5000',))

results = query_env.query(
    '#weight( 0.70 obama 0.20 family 0.10 tree )',
    document_set=map(
        operator.itemgetter(1),
        index.document_ids([
            'clueweb09-en0003-55-31884',
            'clueweb09-en0006-21-20387',
            'clueweb09-enwp01-75-20596',
            'clueweb09-enwp00-64-03709',
            'clueweb09-en0005-76-03988'
        ])),
    results_requested=3,
    include_snippets=True)

for int_doc_id, score, snippet in results:
    # Do stuff with the document and snippet.
\end{minted}
\caption{Advanced querying of topic \texttt{wt09-1} with custom smoothing rules, using a weighted-QLM. Only a subset of documents is searched and we impose a limit on the size of the returned list. In addition to the document identifiers and their retrieval score, the function now returns snippets of the documents where the query terms match.\label{example:advanced_query}}
\end{algorithm}
Beyond simple terms, the \texttt{query()} function fully supports the Indri Query Language.\footnote{\url{http://lemurproject.org/lemur/IndriQueryLanguage.php}}

In addition, we can specify a subset of documents to query, the number of requested results and whether or not snippets should be returned. In Code snippet~\ref{example:advanced_query} we create a \texttt{QueryEnvironment}, with a set of custom smoothing rules. This allows the user to apply fine-grained smoothing settings (i.e., per-field granularity).
\section{Conclusions}

In this paper we introduced pyndri, a Python interface to the Indri search engine. Pyndri allows researchers to access tokenized documents from Indri using a convenient Python interface. By relying on Indri for tokenization and normalization, IR researchers are no longer burdened by this task. In addition, complex retrieval models can easily be implemented by constructing them in the Indri Query Language in Python and querying the index. This will make it easier for researchers to release their code, as Python is designed to be readable and cross-platform. We hope that with the release of pyndri, we will stimulate \textbf{reproducible}, \textbf{open} and \textbf{fast-paced} IR research. More information regarding the available API and installation instructions can be found on Github.\footnote{\url{https://github.com/cvangysel/pyndri}}

\medskip\noindent%
\textbf{Acknowledgements.} This research was supported by the Google Faculty Research Award program and the Bloomberg Research Grant program. All content represents the opinion of the authors, which is not necessarily shared or endorsed by their respective employers and/or sponsors.

\setlength{\bibsep}{0pt}
\bibliographystyle{splncsnat}
{
 \small
 \bibliography{ecir2017-pyndri}
}

\end{document}